\documentclass[useAMS,fleqn,usenatbib]{mnras}

\usepackage{newtxtext,newtxmath}

\usepackage[T1]{fontenc}

\DeclareRobustCommand{\VAN}[3]{#2}
\let\VANthebibliography\thebibliography
\def\thebibliography{\DeclareRobustCommand{\VAN}[3]{##3}\VANthebibliography}

\usepackage{graphicx}	
\usepackage{amsmath}	
\usepackage[T2A]{fontenc}
\usepackage[utf8]{inputenc}
\usepackage[english]{babel}
\usepackage[normalem]{ulem}  
\usepackage{xcolor}
\usepackage[normalem]{ulem}
\usepackage{cancel}

\title[Thermal conduction in the heliosheath]{The strong effect of electron thermal conduction on the global structure of the heliosphere }

\author[V.V. Izmodenov]{
V. V. Izmodenov,$^{1,\,2,\,3}$\thanks{E-mail: izmod@iki.rssi.ru}
D. B. Alexashov$^{1,\,4}$\thanks{E-mail: alexash@ipmnet.ru}
\\
$^{1}$Space Research Institute (IKI) of Russian Academy of Sciences, Moscow, Russia\\
$^{2}$Lomonosov Moscow State University, Moscow center for fundamental and applied mathematics, Moscow, Russia\\
$^{3}$HSE University, 20 Myasnitskaya Ulitsa, Moscow 101000, Russia\\
$^4$Institute for Problems in Mechanics, Russian Academy of Sciences,
Vernadsky Ave. 101, Moscow 117527, Russia
}

\date{Accepted XXX. Received YYY; in original form ZZZ}

\pubyear{2022}

\begin{document}
\label{firstpage}
\pagerange{\pageref{firstpage}--\pageref{lastpage}}
\maketitle

\begin{abstract}
Voyager 1 and 2 crossed the heliopause at $\sim$122 AU in 2012 and $\sim$119 AU in 2018, respectively. It was quite a surprise because the thickness of the inner heliosheath obtained by the  existing at that time models of the global heliosphere  was significantly larger (by 20-40 AU).  Until now, the problem of the heliosheath thickness has not been fully resolved. Earlier in the frame of an oversimplified toy model of nearly isothermal solar wind plasma it has been shown that the effect of electron thermal conduction may significantly reduce the thickness of the inner heliosheath. 
In this paper, we present the first results of  our 3D kinetic-MHD model of the global heliosphere, where the effect of thermal electron conduction has been considered rigorously. The thermal conduction acts mainly along the magnetic field lines. Classical and saturated thermal fluxes are employed when appropriate. \textcolor{red}{ }
It is shown the effects of thermal conduction are significant.  The thickness of the inner heliospheric is reduced. It is desired effect since it helps to reconcile the thickness obtained in the model with Voyager data. The other effects are the strong depletion of the heliosheath plasma temperature toward the heliopause and  the increase of the plasma temperature in the supersonic solar wind upstream of the termination shock.
\end{abstract}

\begin{keywords}
Solar wind --
                Sun: heliosphere --
                Methods: numerical
\end{keywords}



\section{Introduction}

\label{intro}

Voyager 1 (V1) crossed the heliospheric termination shock (TS) at 94 AU in 2004 and the heliopause (HP) at 122 AU in August 2012 \citep{Stone2013, Krimigis2013}.
The V1 crossing of the heliopause at 122 AU was not expected by a part of the heliospheric community since the global models of the solar wind (SW) - local interstellar medium (LISM) interaction suggest that the thickness of the inner heliosheath, which is the region between the TS and the HP, in the V1 direction should be of the order of 50-70 AU depending on the model \citep[see, e.g.,][]{Izmodenov2013}.

The Voyager 2 crossings of the  heliospheric termination shock at 84 AU in 2007 \citep{Decker2008, Stone2008} and the heliopause at 119 AU in 2018 \citep{Burlaga2019, Gurnett2019, Krimigis2019, Richardson2019, Stone2019} are generally confirmed that the measured thickness of the inner heliosheath is smaller than expected in the models. For example, in the time-dependent model of \cite{Izmodenov2020} the heliopause distances  are $\sim$123.5 AU in August 2012 for Voyager 1 direction and $\sim$121.5 AU in November 2018 for Voyager 2 direction that is quite close to the actual crossings. However, the distances to the termination shock obtained in the model by 14 and 6 AU closer than actual distances of crossings.
It was also shown in that paper that the angle between interstellar magnetic field and LISM velocity has strong influence on locations of the termination shock and the heliopause. However, it is not possible to match the Voyager's TS and HP crossings by varying the magnitude and direction of the magnetic field.

After 10 years since 2012, the problem of the heliosheath thickness has no definitive solution. Several ideas to resolve this problem appeared recently in the literature. \cite{Borovikov2014} suggested that the smaller distance is connected with the instabilities of the heliopause. \cite{Schwadron2013} argued for the interstellar flux transfer effect. \cite{Swisdak2013} and \cite{Opher2013} explained the observed behavior by magnetic reconnection in the inner heliosheath and at the heliopause.

The importance of  electron thermal conduction in the heliosheath plasma flow has been studied by \cite{Izmodenov2014} in the frame of an over-simplified 'toy' model by assuming that the plasma flow is isothermal in the entire heliosphere.  The toy model is clearly demonstrated that the effect of thermal conduction may be significant, and it strongly reduces the thickness of the inner heliosheath, which helps to reconcile the model results with Voyager observations. 

This paper explores the effect of thermal electron conduction. We present the first results of the 3D kinetic-MHD model of the solar wind interaction with the local interstellar medium in which the effect of thermal conduction has been taken into account.

\section{Model}\label{gov_eqns}

In the presented model, we use a single fluid ideal MHD approach for all charged particles (electrons, protons, pickup protons, helium ions).
The system of governing equations is the following:

\begin{equation}
\frac{\partial\rho}{\partial t}+\nabla\cdot(\rho\boldsymbol{V})=Q_n, 
\label{eq1}
\end{equation}

\begin{equation}
\frac{\partial(\rho\boldsymbol{V})}{\partial t}+\nabla\cdot\left[\rho{\boldsymbol{V}}{\boldsymbol{V}} + \left(p + \frac{B^2}
{8\pi}\right) I - \frac{{\boldsymbol{B}}{\boldsymbol{B}}}
{4\pi}\right] =
 \boldsymbol{Q}_i,
\label{eq2}
\end{equation}

\begin{equation}
\frac{\partial E}{\partial t}+\nabla\cdot\left[\left(E + p + \frac{B^2}{8\pi}\right){\boldsymbol{V}} -
\frac{(\boldsymbol{V}\cdot\boldsymbol{B})}{4\pi}\boldsymbol{B}\right] =
 Q_e + \rm{div} \boldsymbol{q},
\label{eq3}
\end{equation}

\begin{equation}
\frac{\partial\boldsymbol{B}}{\partial t}+\nabla\cdot({\boldsymbol{V}}{\boldsymbol{B}} - {\boldsymbol{B}}
{\boldsymbol{V}})=0, \qquad \mbox {div} \, {\boldsymbol{B}} = 0
\label{eq4}
\end{equation}

Here $\rho$, $\boldsymbol{V}$, and $p$ are the density, velocity, and pressure of the plasma component, respectively.
 $\boldsymbol{B}$ is the magnetic field induction vector, $E = \frac{\rho V^2}{2}+\frac{p}{\gamma-1}+\frac{B^2}{8\pi}$ is the total energy, $\gamma = 5/3$.  The sources $Q_n, \boldsymbol{Q}_i, Q_e$ in the right parts of (\ref{eq1}-\ref{eq3})  are due to processes of charge exchange with neutral component and photoionization. 

The system of equations (\ref{eq1}-\ref{eq4}) is the same as in the models presented by \cite{Izmodenov2015, Izmodenov2020} with only one exception of the term $\rm{div} \boldsymbol{q}$.

The system of equations (\ref{eq1}-\ref{eq4}) is solved together with the kinetic equation for the neutral component consisting of hydrogen atoms. This equation, together with expressions for the source terms  $Q_n, \boldsymbol{Q}_i, Q_e$, can be found in \cite{Izmodenov2015}. We do not repeat them here.

The exclusive goal of this paper is to establish the effect of the thermal conduction term $\rm{div} \boldsymbol{q}$ in the energy equation.
The influence is analyzed by comparing the modeling results with the model without thermal conduction. For such a model we choose Model 1 from \cite{Izmodenov2020}.  The boundary conditions for this model are also presented in \cite{Izmodenov2020}.
The boundary conditions use the solar wind data at 1 AU and the interstellar parameters inferred from Ulysses/GAS data and other previous studies. For the less known interstellar parameter that is interstellar magnetic field (IsMF) we choose $B_{LISM} = 3.75$ $\mu$Gauss and the direction that lies within the Hydrogen Deflection Plane \citep[see,][]{Lallement2005, Lallement2010} and constitutes $60^{\circ}$  with the direction of interstellar flow vector.

\subsection{Thermal conduction}

According to the classical theory of \cite{Spitzer1962}, the heat conduction vector is determined as follows

\begin{equation}
\boldsymbol{q} = \kappa T_e^{5/2} (\boldsymbol{b}\cdot \nabla T_e) \boldsymbol{b},\;\; \boldsymbol{b}=\boldsymbol{B}/B.
\label{eq5}
\end{equation}

As it is seen, the thermal flux is not isotropic. It is maximal in the direction of the magnetic field and reduced in the perpendicular direction.  The coefficient for thermal conductivity $\kappa$ is expressed through the Coulomb logarithm 
\citep[][]{Balbus1986}
\[
\kappa =\frac{1.84 \cdot 10^{-5}}{ln \Lambda} ergs\; cm^{-1} s^{-1} K^{-7/2}, 
\]

where $ln \Lambda = 29.7 + ln[T_e/(10^6 K) (n_e/(1 cm^{-3}))^{-1/2}]$.

In the considered problem, the Coulomb logarithm varies in the range
 $ln \Lambda \approx 27 - 33$ 
 (29-33 in supersonic solar wind and heliosheath, 27-31 in interstellar medium, and 29-31 close to heliopause).
For simplicity, we used averaged value 30 and corresponding coefficient  
 $\kappa \approx 6\cdot 10^{-7} (in\, sgs)$.

The classical formula above is valid under following condition

\begin{equation}
\lambda 
\ll \frac{T_e}{|\nabla T_e|}.
\label{eq6}
\end{equation}
where $\lambda$ is the electron mean free path  with respect to electron collisions. 
This means that the electron mean free path should be much smaller than characteristic scale of the temperature variation.

In our problem, the condition (\ref{eq6}) does not work at the fronts of the shocks and heliopause.
Moreover, it does not work also in the inner heliosheath between the heliospheric termination shock and the heliopause. In this case the thermal flux is so-called saturated and can be calculated as follows \citep{Cowie1977}
\begin{eqnarray}
   \mathbf{q} = q_{sat} sgn (\boldsymbol{b}\cdot \nabla T_e) \boldsymbol{b}, \label{eq7} \\
    q_{sat} = 0.4\left(\frac{2k_b T_e}{\pi m_e}\right)^{1/2} n_e k_b T_e =
5 \phi \rho (p/\rho)^{3/2}. \label{eq8}
\end{eqnarray}
Here $n_e$ is the electron number density, $m_e$ is the electron mass, $k_b$ is the Boltzmann constant.
$\phi$ is a constant that can be chosen in the range of $[0,1.1]$ depending on the ionization state of the gas.
In the case of fully ionized plasma in thermal equilibrium  with $\phi \approx 1$. 
In our model  $\phi = 0.5$ is assumed since the neutral component is present. We also performed calculations with different values of $\phi$.

In the numerical calculations for each cell of the numerical grid  thermal fluxes are calculated by (\ref{eq5}) and (\ref{eq7}). Following \cite{Cowie1977}, we choose classical flux (\ref{eq5}) when the classical flux is smaller than the saturated one (\ref{eq7}), and the saturated flux otherwise.


Following \citet{Izmodenov2020}, the helium ions in the interstellar medium and solar alpha particles have been taken into account in the model  \citep[see, also,][]{Izmodenov2003}. In this case, 
to close the system of equations the following equation of state  is employed: $p = (n_p + n_{He} + n_e) k_b T$,
where $n_p$, $n_e$ and $n_{He}$ are the number densities of protons, electrons and helium ions, From conditions of quasi-neutrality:  $n_e = n_p + 2 n_{He}$ in the solar wind, $n_e = n_p + n_{He}$ in the  interstellar medium. In this notation,  
$\rho = m_p (n_p + 4 n_{He})$, $m_p$ is the proton mass. 
Introducing the parameter $\delta$ that is
$\delta = (n_p + n_{He})/(n_p + 4 n_{He})$ for the interstellar medium and
$\delta = (n_p + 1.5 n_{He})/(n_p + 4 n_{He})$ for the solar wind,
we have the following expressions for the electron temperature and the saturated heat flux $q_{sat}$:
\[
T_e = T = \frac{m_p}{2 k_b} \frac{p}{\delta \rho},\;\;
q_{sat} = 5 \phi m_p n_e \left(\frac{p}{\delta\rho}\right)^{3/2}.
\]

\section{Numerical approach}

The problem to be solved numerically consists in the solution of system of equations (\ref{eq1}) - (\ref{eq4}) self-consistently with kinetic equation for the neutral component. The latter equation is solved using the Monte-Carlo method (\cite{Malama1991}, see, also,  \cite{Izmodenov2015}).

The system (\ref{eq1}) - (\ref{eq4}) is solved by Yanenko's method of fractional steps \citep{Yanenko}.
In this numerical approach, the solution of the full system on the next time step is obtained by two (fractional) sub-steps. In the first sub-step, we solve 
the thermal conduction equation:
\begin{equation}
    \frac{\partial \varepsilon}{\partial t} =  \rm{div} \boldsymbol{q},
    \label{eq_term_cond}
\end{equation}
where $ \varepsilon = \frac{p}{\gamma-1}$ is the plasma thermal energy.
As a result of this solution new distribution of the thermal pressure is obtained.  This distribution is used as initial values of pressure
for the second sub-step. In the second sub-step, we solve the ideal MHD equations with the source terms due to charge exchange.  The numerical procedure for this time-step is the same as described before in \cite{Izmodenov2015, Izmodenov2020}. Briefly, to solve the ideal MHD equations we use finite-volume high-order Godunov's type scheme that includes 3D adaptive moving grid with discontinuities capturing and fitting capabilities, Harten-Lax-van Leer Discontinuity (HLLD) MHD Riemann solver and Chakravarthy-Osher TVD procedure. 


The numerical methods employed for the first time-step are described below. To calculate the thermal flux, we preliminary construct the magnetic field lines passing through the centers of computational cells. The magnetic field lines are calculated for all computational cells and at each computational time-step. Then the thermal fluxes are calculated by solving a one-dimensional version of the thermal conduction equation (\ref{eq_term_cond}) along of each field line:
\begin{equation}
    \frac{\partial \varepsilon}{\partial t} = \frac{\partial q}{\partial s} - \frac{q}{B} \frac{\partial B}{\partial s} = B \frac{\partial}{\partial s}\left(\frac{q}{B}\right),
    \label{eq_term_cond_2}
\end{equation}
where $s$ is the coordinate along the magnetic field line, $q = |\mathbf{q}|$, $B = |\mathbf{B}|$. Note, the equation (\ref{eq_term_cond_2}) has been derived from (\ref{eq_term_cond}) under assumption that 
the vectors $\mathbf{q}$ and $\mathbf{B}$ are parallel. 



The equation (\ref{eq_term_cond_2}) is solved by implicit second order numerical scheme:
\begin{equation}
 \frac{ \varepsilon^{n*}_i - \varepsilon^{n}_i}{t^{n+1}-t^{n}} = \frac{B_i}{h}\left[
 \frac{ q_{i+\frac{1}{2}}^{n*} }{B_{i+\frac{1}{2}}}
-\frac{q_{i-\frac{1}{2}}^{n*}}{B_{i-\frac{1}{2}}}\right],
     \label{eq_term_cond_scheme}
\end{equation} 
where $t^{n}$ and $t^{n+1}$ are the time moments corresponding to previous and following time steps, $h$ is the uniform step along the magnetic field line.
For the classical thermal flux $q_{i\pm\frac{1}{2}}^{n*}$ is approximated as
\begin{equation}
        (q_{i\pm\frac{1}{2}}^{n*})_{cl}= \pm 
        \kappa \left(\frac{T_{i\pm 1}+T_i }  
        {2}\right)^{5/2} \frac{T_{i\pm 1}-T_i}{h},\;\;  
     \label{eq_qh_classic}
\end{equation}
where
\begin{equation}
        T_{i\pm 1}=\frac{1}{k_b}\frac{p_{i\pm1}^{n*}}{N_{i\pm1}^{n}},
        \;\;  T_i = \frac{1}{k_b}\frac{p_{i  }^{n*}}{N_{i  }^{n}}, \nonumber
\end{equation}
and $N$ is sum of number densities of protons, electrons 
and helium ions.
For the saturated flux  $q_{i\pm\frac{1}{2}}^{n*}$ is calculated as
\begin{equation}
 (q_{i\pm\frac{1}{2}}^{n*})_{sat} = \pm 
 5\phi (n_e)_{i\pm\frac{1}{2}}^{n} 
        \frac{(2k_b)^{3/2}}{\sqrt{m_p}} \left(\frac{T_{i\pm 1}+T_i }   
{2}\right)^{3/2}  
        \frac{\Delta T_{i \pm 1} }{|\Delta T_{i \pm 1}|},
     \label{eq_qh_saturated}
\end{equation}
where $\Delta T_{i \pm 1} =  T_{i \pm 1} - T_i$.

The choice between classic and saturated fluxes depends on 
their absolute values. Eventually, we have
$q_{i\pm\frac{1}{2}}^{n*}= (q_{i\pm\frac{1}{2}}^{n*})_{cl}$ if
$|(q_{i\pm\frac{1}{2}}^{n*})_{cl}|<|(q_{i\pm\frac{1}{2}}^{n*})_{sat}|$ and
$q_{i\pm\frac{1}{2}}^{n*}= (q_{i\pm\frac{1}{2}}^{n*})_{sat}$ overwise.

To get the values of $p^n_i$ and $\rho^n_i$ at the points along the magnetic streamline,
we use a linear interpolation procedure.

To solve non-linear system of equations (\ref{eq_term_cond_scheme}) for  intermediate values $\varepsilon^{n*}_i$
we used an iterative procedure. 


Solving the system of non-linear equations (\ref{eq_term_cond_scheme}) by an iterative procedure, we obtained the pressure distribution that is used then to calculate mass, momentum and energy fluxes at the time-step of Godunov's method.

\begin{figure}
\begin{center}
\includegraphics[width=1.0\linewidth]{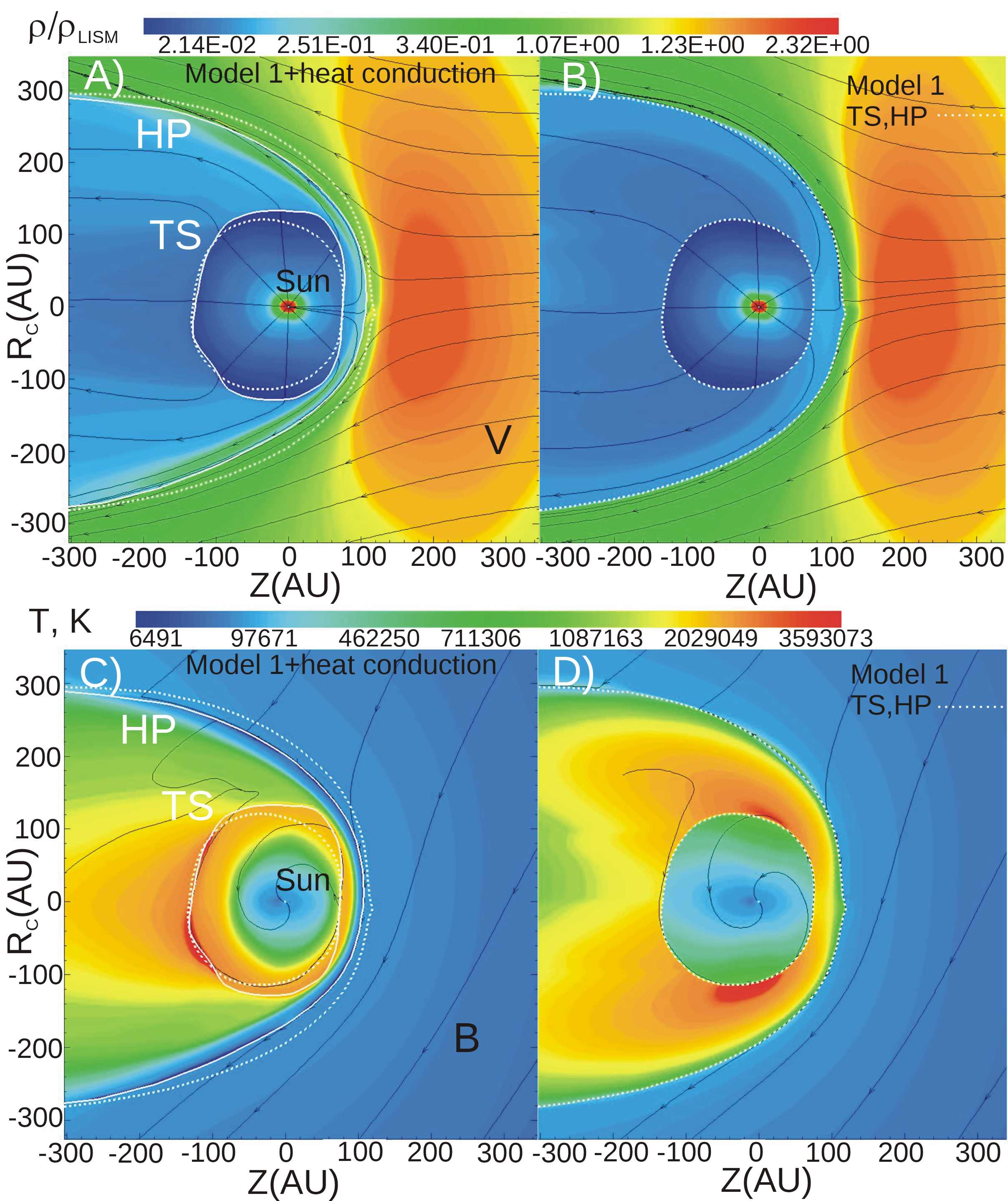}
\caption{ Number density (panels A and B) and temperature (panels C and D) of the solar wind in the solar polar plane. Panels A and C present the results obtained in the frame of the model with the effect of thermal conduction. Panels B and D are shown for the sake of comparison and present result of Model 1 of \protect\cite{Izmodenov2020}. The shapes of the heliospheric termination shock and the heliopause are shown as the white curves. Solid and dashed lines correspond to the current model and Model 1, respectively. Projections of the streamlines and magnetic field lines are shown as black lines.}
\label{fig1}
\end{center}
\end{figure}

\begin{figure}
\begin{center}
\includegraphics[width=1.0\linewidth]{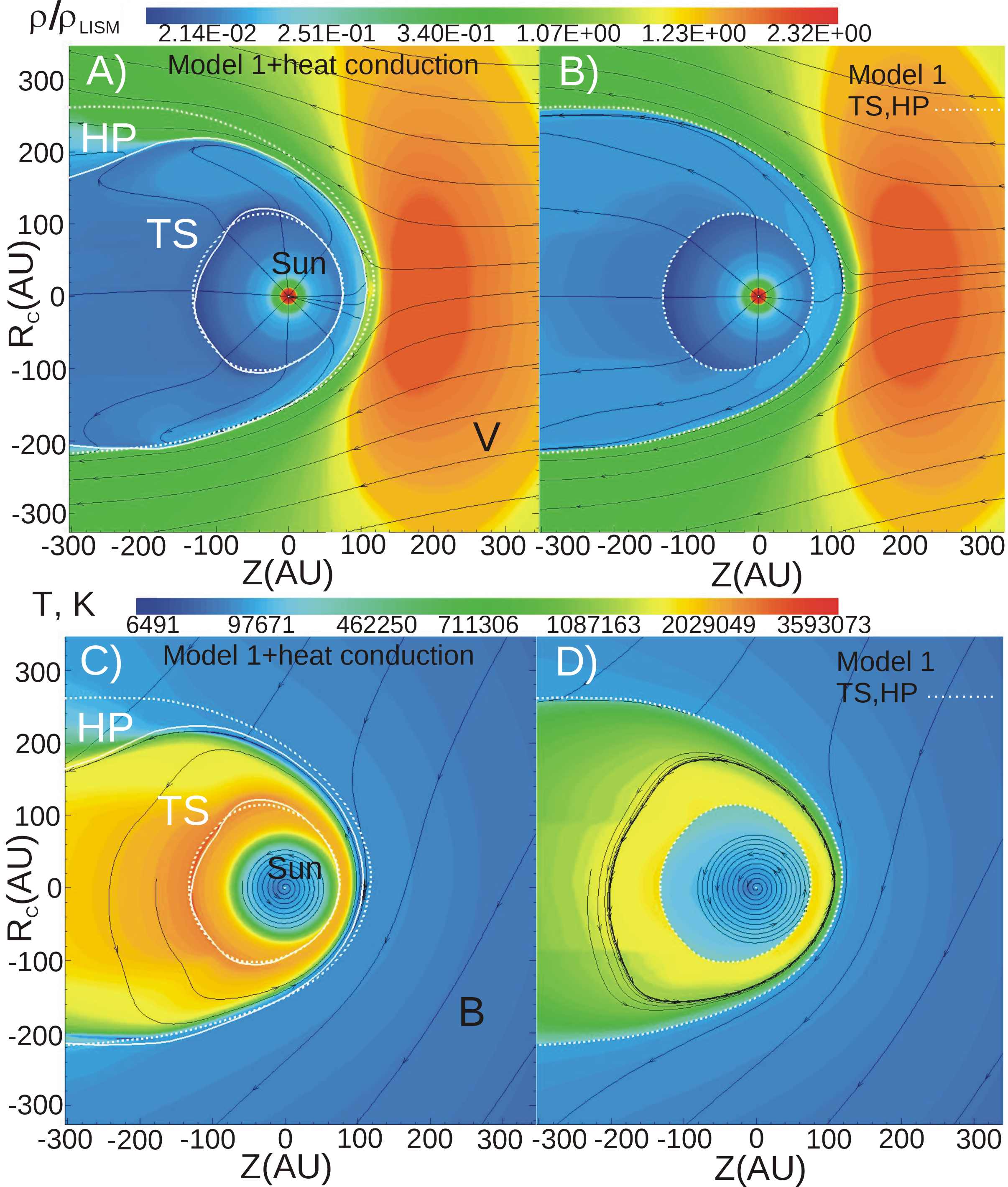}
\caption{Description is  the same as for Figure 1, but for equatorial plane.}
\label{fig2}
\end{center}
\end{figure}

\begin{figure}
\begin{center}
\includegraphics[width=1.0\linewidth]{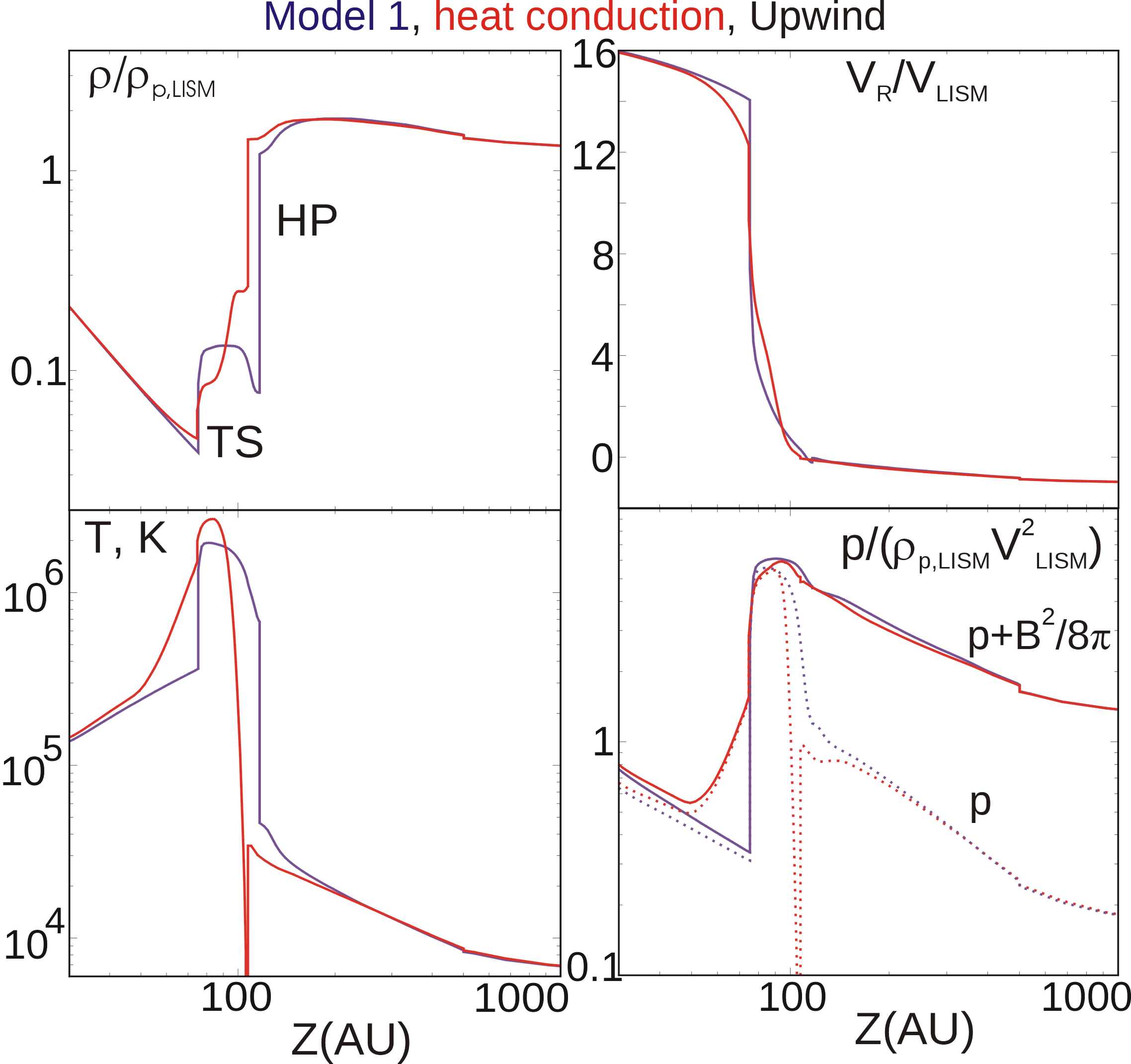}
\caption{ Density ($\rho$), radial component of velocity ($V_R$), temperature  ($T$) and total ($p+ \frac{B^2}{8 \pi}$) pressure   of the plasma component as functions of the heliocentric distance. The distributions are shown in the upwind direction. Red curves correspond to the results obtained in the frame of current model with thermal conduction included. Black curves correspond to Model 1. }
\label{fig3}
\end{center}
\end{figure}

\begin{figure}
\begin{center}
\includegraphics[width=1.0\linewidth]{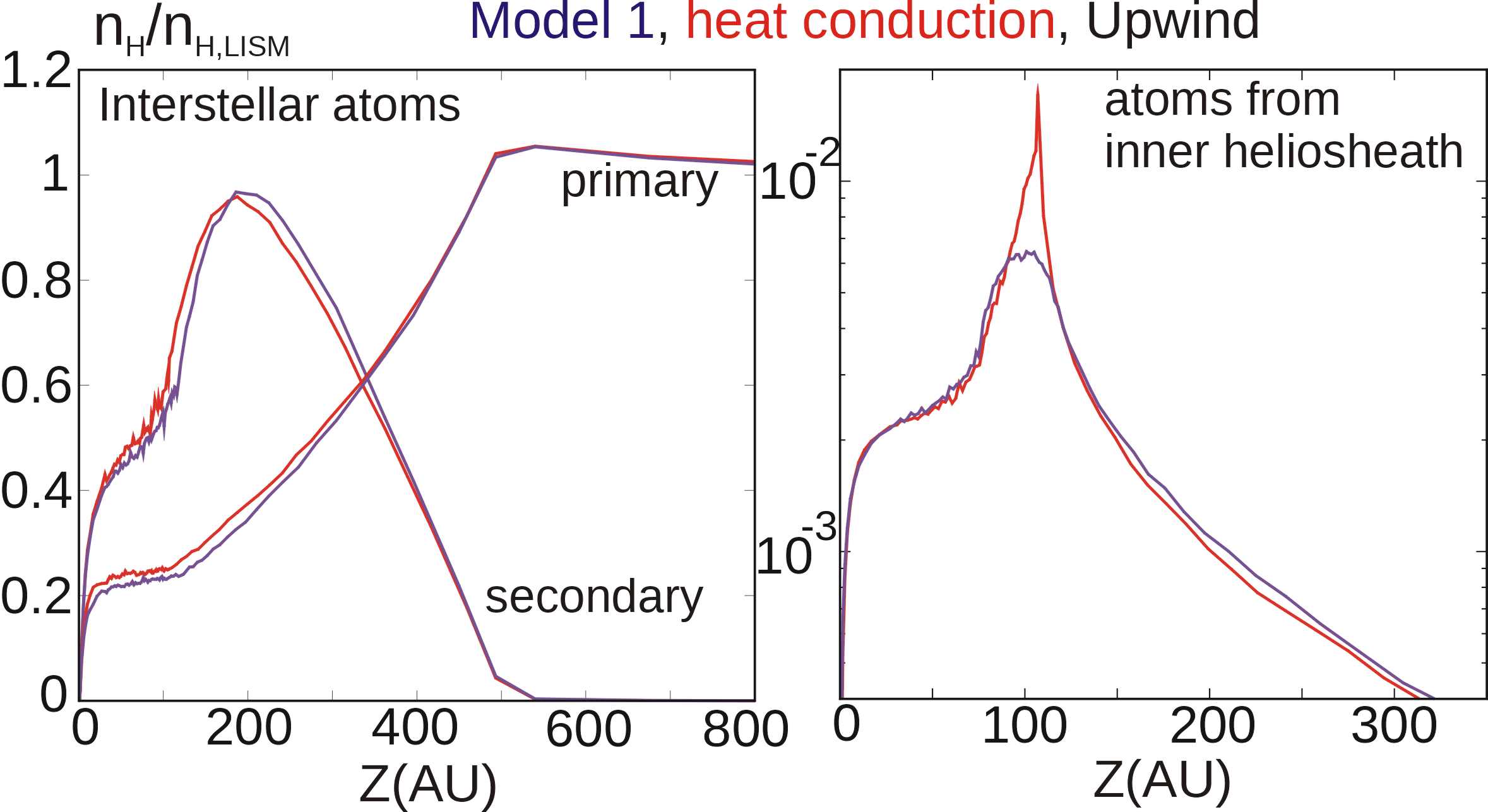}
\caption{ Number densities of different populations of atomic hydrogen as functions of the heliocentric distance. The distributions are shown in the upwind direction.
Red curves correspond to the results obtained in the frame of current model with thermal conduction included. Black curves correspond to Model 1.}
\label{fig4}
\end{center}
\end{figure}


\section{Results}
\label{results}

The effect of thermal conduction on the global heliosphere is significant. 
It is clearly seen from Figures \ref{fig1} and \ref{fig2}. The figures present the results in the solar polar and equatorial planes, respectively. The comparison of the current model results with the results of the model without thermal conduction (Model 1, hereafter) is presented.
The most pronounced effect is in the inner heliosheath between the TS and HP. The temperature in this region is high  ($T \gtrsim 10^6$ K), so  thermal fluxes are large and saturated.

The results show that the thickness of the heliosheath is reduced by about  20 \% in the region of the solar equator mainly due to the heliopause moving toward the Sun (compared to Model 1). At the poles the heliosheath thickness is reduced by more than 50 \%.
In addition to the HP that approaches the Sun, the TS in the pole regions moves out (compared to Model 1).
 This happens due to higher solar wind speed at the poles, which results in higher temperature in the post-shocked plasma of the inner heliosheath. The higher temperature, the larger the coefficient of the heat conduction.

The changes in shapes and locations of the heliopause and termination shock are connected with strong depletion of the plasma temperature and pressure in the inner heliosheath toward the heliopause (Figure \ref{fig3}).
The plasma temperature drops in the vicinity of the heliopause so-strongly that it becomes much smaller than the temperature from the interstellar side of the heliopause.
This is the combined effect of the thermal conduction and magnetic field. Indeed, after crossing the TS, the plasma temperature is high, so thermal conduction works effectively. The heat is removed from the region toward the tail along the magnetic field lines. Therefore, the temperature reduces as we move from the TS toward the heliopause. The coefficient of thermal conduction becomes small, and, at some distances the thermal conduction becomes ineffective. 

Moving further toward the heliopause, the observer enters the so-called magnetic wall region, where the magnetic field becomes dominant. The plasma moves around this region, creating plasma depletion.
The effect has been obtained and explained in  \cite{Izmodenov2015}. It is also seen in Figure \ref{fig3} for Model 1. The temperature also drops down in the vicinity of the termination shock. Nevertheless, the temperature of the heliospheric plasma at the heliopause is about 10 times higher than the interstellar temperature.

For the current model, the effects of thermal conduction and plasma depletion due to magnetic field work simultaneously. This produces huge drop of heliosheath temperature at the heliopause.

This is due to the fact that the thermal conduction works along the magnetic field lines, which in the upwind, are parallel to the HP surface and nearly parallel to the TS. The magnetic field lines connect the heated plasma of the nose region with cooler plasma in the downwind. The thermal conduction redistributes heat toward the tail. This mechanism is more effective for the streamlines closer to the heliopause since they are connected to the most distant tail regions, which are cooler.

In the outer heliosheath beyond the heliopause, the influence of thermal conduction is much less pronounced, and it is noticeable only in the vicinity of the heliopause. 
 
There is another interesting effect that is seen in  Figures \ref{fig1} - \ref{fig3}. The plasma temperature in the supersonic solar wind upstream of the TS is lager compared with Model 1. This is due to transfer of thermal energy from the heliosheath due to thermal conduction along the magnetic steamlines. The streamlines connect the supersonic solar wind region with the inner heliosheath because the TS is not spherical. This effect is rather large and will be discussed in last section.
 
It is interesting to note that the termination shock remains to be discontinuity despite the presence of a dissipating process of the thermal conduction. It can be explained by the form of the saturated thermal flux expression employed. This flux (contrary to the classical flux) does not change the divergent form of governing equations. Therefore, the Rankine-Hugoniout (R-H) conditions connect the parameters upstream and downstream the shock. The only difference with classical case is the appearance of the additional (algebraic) term in the R-H condition.
 Such modification of the R-H conditions is well known (see, for example, \cite{Chernyi1988}). Therefore, the shock structures remain for thermal conduction with the saturated heat flux. For the classical heat flux, one could expect a smooth shock transition when the magnetic field streamline is not parallel to the shock front. If the streamline is nearly parallel to the shock front then the transition region is quite narrow.

Figure \ref{fig4} presents the number density distribution of the neutral components in the upwind direction.
The left panel presents the number density of interstellar atoms (both primary and secondary) in the upwind direction.

The difference with Model 1 is within  5 \%.
A much more difference with Model 1 is seen in the number density of H atoms, which were born (by charge exchange with protons). The number density in the current model is larger by a factor of two at the maximum (see the right panel in Figure \ref{fig4}). This is connected with the significant increase in the number density of the parent protons since the charge exchange rate is proportional to the density.

\section{Summary and discussion}
\label{sum}

The presented first results of the global model of the heliosphere confirm the preliminary estimates of the importance of thermal conduction on the plasma flow in the inner heliosheath. The thermal conduction essentially changes the shapes of the heliopause and the termination shock. The thickness of the inner heliospheric is significantly (by a factor of two at the poles) reduced. This is the desired effect because it helps to reconcile the thickness obtained in the model with Voyager data.

Other main results obtained are connected with plasma temperature in the inner heliosheath and the supersonic solar wind.
The first not evident and expected result is the strong depletion of the heliosheath plasma temperature toward the heliopause.
The depletion is so strong that the temperature becomes lower than the interstellar temperature. We explain the depletion by combined effects of the thermal conduction and magnetic field. 

The second result is the increase of the plasma temperature in the supersonic solar wind upstream of the termination shock.
This increase is associated with the heat transfer from the inner heliosheath due to the thermal conduсtion.

Due to the lack of electron temperature observations in the distant heliosphere and the heliosheath, we can not verify this result directly. However, a single-fluid approach for plasma is assumed in this model. Such an assumption has been employed in our previous models \citep[except for][] {Malama2006,  Chalov2016, Baliukin2020, Baliukin2022}.  It is implied that the pickup protons 
immediately assimilate into the solar wind plasma. This leads to a large increase in the proton temperature toward the termination shock and, therefore, the electron temperature.  Such an increase was not observed by Voyager 2, which proved the thermal decoupling of the solar protons and pickup protons. Nevertheless, the non-adiabatic behaviour of the solar wind protons is observed \citep{Gazis1994, Lazarus1995} and was explained by the transfer of thermal energy from pickups to protons through their interaction with electromagnetic waves propagating from the Sun and originating due to unstable ring distribution of newly originated pickup protons \citep{Williams1995}. Recently \cite{Korolkov2022} have shown that the primary process responsible for the non-adiabatic behaviour of the distant solar wind is the heating by numerous shock waves propagating in the solar wind.

The single-component approach for the heliospheric plasma is probably appropriate to describe the global structure of the heliosphere since it is built on fundamental laws of mass, momentum, and energy conservation. Nevertheless, describing distribution of different components requires a multi-component approach as those proposed by  \cite{Malama2006}, \cite{Chalov2013}, and later by \cite{Chalov2016}, \cite{Baliukin2020, Baliukin2022}.
In addition to separation of solar and pickup protons, most probably, the electron component should be considered as thermally decoupled with solar protons. Such a model has been recently developed for the inner heliosphere by \cite{Usmanov2018}. Such an approach should be extended to the global models of the heliosphere. The possible extension will require the conditions for solar protons, pickup protons and electrons at shocks in general, and, in particularly, at the termination shock. Such conditions have been considered by 
\cite{FahrChalov2008}, \cite{Fahr_etal2012}, \cite{FahrSiewert2013}, \cite{FahrSiewert2015} under assumption of magnetic moment conservation across the shock. Specific role of electrons has been explored in \cite{Fahr_etal2015}. Recently, \cite{Gedalin_etal2021} obtained conditions for downstream pickup proton temperature and pressure by using test-particle analyses.

\begin{figure}
\begin{center}
\includegraphics[width=1.0\linewidth]{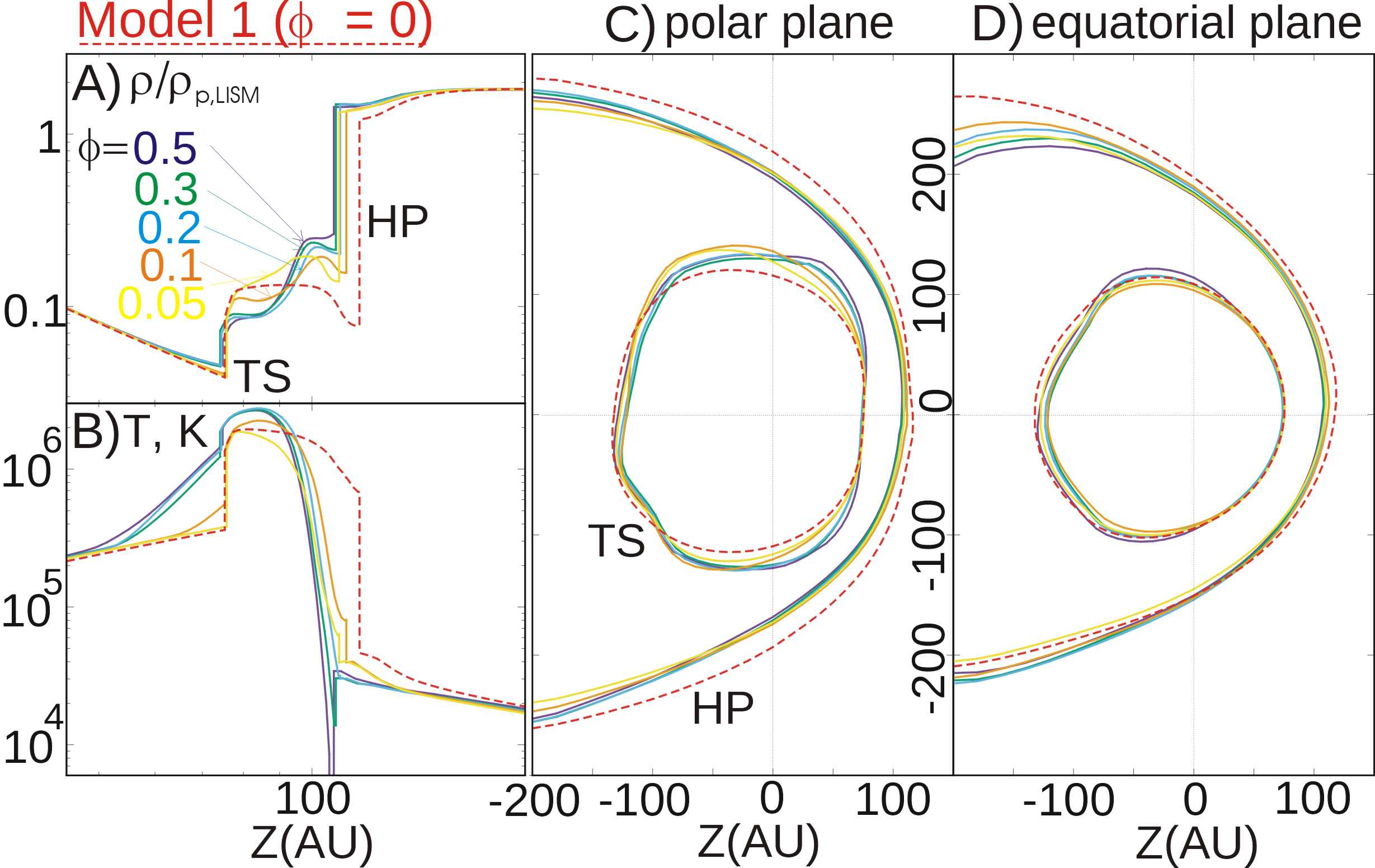}
\caption{ The distribution of number density (A) and temperature (B) in the upwind direction. Different curves correspond to the different values of the parameter $\phi$ of equation (\ref{eq8}). Panels C 
 and D shows the positions of the termination shock and the heliopause in the polar and equatorial planes. }

\label{fig_phi}
\end{center}
\end{figure}

Another way to advance the current model of the global heliosphere is to consider the parameter $\phi$ in the expression (\ref{eq8}) as a function of coordinates. Indeed, this coefficient depends on the  ionization state ($\phi =1$ for fully ionized plasma), which strongly changes throughout the heliosphere.
The presented solution strongly depends on $\phi$. To demonstrate it, we performed calculations for different values of $\phi$ (see, Figure \ref{fig_phi}).

It is also important to underline that in the considered model the thickness of the inner heliosheath is reduced due to evacuation of thermal energy into the heliotail along of magnetic field lines. The 
 the inner heliosheath would be reduced if the heat is transferred through the heliopause. However, the electron heat conduction is strongly anisotropic and the heat is transferred along the magnetic field lines by several orders of magnitude faster than in the perpendicular directions. The magnetic field lines are parallel to the heliopause surface. Therefore, despite much longer way along the magnetic field line the heat energy will evacuated in the heliotail region rather than transferred through the heliopause.  This approach is considered in our paper.
Of course, alternatively, some dissipative process may lead to reconnection of  the magnetic field lines at the heliopause. Then the thermal energy can be evacuated through the reconnected lines.

Overall, we conclude that thermal conduction has a significant effect on the global shape of the heliosphere and should be included in the global models. At the same time, there is no doubt that single fluid description is not appropriate to be used for comparison with Voyager plasma data. Instead, most probably, pickups, solar protons, and electrons should be considered as three co-moving but thermally decoupled fluids.

\section*{Acknowledgements}

The work was performed in the frame of the Russian Science Foundation grant 19-12-00383.
We thank our reviewer, Hans Fahr, for very fruitful comments and  Igor Baliukin for help with preparation of the paper. 

\section*{Data Availability}
 

The  data underlying this article will be shared on reasonable request to the corresponding author.

\appendix

\bsp	
\label{lastpage}
\end{document}